# Discrimination and AI in insurance: what do people find fair? Results from a survey

Frederik Zuiderveen Borgesius[a], Marvin van Bekkum[a], Gabi Schaap, Iris van Ooijen, Maaike Harbers Tjerk Timan

Two modern trends in insurance are data-intensive underwriting and behavior-based insurance. Data-intensive underwriting means that insurers analyze more data for estimating the claim cost of a consumer and for determining the premium based on that estimation. Insurers also offer behavior-based insurance. For example, some car insurers use artificial intelligence (AI) to follow the driving behavior of an individual consumer in real-time and decide whether to offer that consumer a discount. In this paper, we report on a survey of the Dutch population (N=999) in which we asked people's opinions about examples of data-intensive underwriting and behavior-based insurance. The main results include: (i) If survey respondents find an insurance practice unfair, they also find the practice unacceptable. (ii) Respondents find almost all modern insurance practices that we described unfair. (iii) Respondents find practices for which they can influence the premium fairer. (iv) If respondents find a certain consumer characteristic illogical for basing the premium on, then respondents find using the characteristic unfair. (v) Respondents find it unfair if an insurer offers an insurance product only to a specific group. (vi) Respondents find it unfair if an insurance practice leads to the poor paying more. We also reflect on the policy implications of the findings.

## 1 INTRODUCTION

Insurers underwrite risks by analyzing data. Insurers increasingly use Artificial Intelligence (AI): based on correlations found in data, the insurer estimates the expected claims cost of the consumer and assigns the consumer to a risk class. Two modern trends in insurance are (i) data-intensive underwriting and (ii) behavior-based insurance. (i) Data-intensive underwriting means that insurers use and analyze more data for risk underwriting. Insurers could extend their analyses with machine learning. For example, insurers could collect more data about consumers and use machine learning to find new correlations in data to underwrite risks with. Some of those correlations could look odd to the consumer, such as the correlation between a consumer's house number and the probability that they file an insurance claim (ii) A second trend is behavior-based insurance: insurers can analyze the behavior of an individual consumer to offer that consumer a discount. For example, some insurers offer a discount to the consumer if that consumer allows the insurer to monitor their driving behavior (behavior-based car insurance). Other insurers offer discounts to health or life insurance when consumers show, with a health tracker, that they have an active lifestyle.[1]

Both modern insurance trends can have advantages. For instance, careful consumers or consumers who run less risk may be able to insure themselves for a lower premium.[2] But both trends may also have negative effects. For example, the insurer could, unintentionally, illegally discriminate certain groups. Other effects could be controversial or unfair, even if the insurer does not discriminate illegally. This paper focuses on such possibly unfair – but not necessarily illegal – effects.

While there is much research discussing these trends, less is known about what the public thinks about data-intensive and behavior-based decisions in the insurance sector. We performed a survey of the Dutch population (N=999) in which we asked respondents for their opinions about examples of (i) data-intensive underwriting and (ii) behavior-based insurance. In general, we wanted to know whether, and under which circumstances, the general public finds the practices fair and acceptable. The paper explores the following questions.

(1) To what extent do people find data-intensive underwriting and behavior-based insurance fair and acceptable?
(2) If people feel they have more influence on the premium, do they find an insurance practice fairer and more acceptable?
(3) If people find an insurance practice more logical, do they find the practice fairer and more acceptable?
(4) If an insurer only accepts certain parts of the population, do people find such practices less fair and acceptable?

---

[a] Joint first author.
[1] See section 2.
[2] However, section 2 shows that the trends may also lead to more expensive insurance for some, and to exclusion.

(5) If an insurance practice leads to higher prices for poorer consumers, do people find the practice less fair and less acceptable?

The paper makes the following contributions to the literature. First, this is one of the first academic papers to research what the public finds fair in the context of modern applications of AI in insurance.[3] This is an important addition to the literature, as fairness is always context-specific.[4] Second, we are the first to ask respondents about specific types of insurance fairness. For example, we explore what respondents think about insurers basing the premium on consumer characteristics that may seem illogical, or irrelevant, to the respondent.

Our results can be useful for several groups of readers. For example, the results could inspire scholars or others interested in perceived fairness of AI and data use, as the survey gives an impression of the attitudes of the average citizen for real-life examples. Such surveys are scarce: the perspectives of insurance consumers are "generally an under-researched area".[5] Second, insurers who use AI in their business, or plan to do so, may want to know what respondents think of modern insurance trends.[6] Third, policymakers who consider regulating AI in insurance could use the survey as a first impression of what the public thinks.[7] We do not claim that policymakers should always prohibit practices that seem unfair according to surveys.[8] The results could be of use in other sectors outside of insurance too, as insurance is an example of a sector that has relied on statistics and AI for a relatively long period of time.[9]

We aim to avoid unnecessary jargon, so that the paper is understandable for readers without a background in insurance or actuarial science. We conducted the survey in Europe, and we discuss some EU law in the discussion section. The paper's results can be useful outside the EU too, for example, by providing ideas for further research in other countries, allowing transnational applications of AI, or by enabling more (public) discussion about the topic.[10]

In Section 2, building on earlier work, we provide background to our survey questions. Section 3 discusses the method for the survey. Section 4 presents the results, with each subsection discussing a different research question. In section 5 we reflect on the results and place them in broader context, such as whether the law and fairness metrics can help against the perceived unfairness. Section 6 discusses limitations of the research and provides suggestions for further research. Section 7 concludes.

---

[3] See for a rare example of another survey about people's opinions on modern insurance practices (in the US, in that paper): Barbara Kiviat, 'Which Data Fairly Differentiate? American Views on the Use of Personal Data in Two Market Settings' (2021) 8 Sociological Science 26 <https://sociologicalscience.com/articles-v8-2-26/>. Binns et al. investigate in experimental studies "people's perceptions of justice in algorithmic decision-making under different scenarios and explanation styles', including in the context of insurance. Reuben Binns et al., 'It's Reducing a Human Being to a Percentage" Perceptions of Justice in Algorithmic Decisions.' Proceedings of the 2018 Chi conference on human factors in computing systems. 2018.  For an overview of empirical literature until 2022 on people's fairness perceptions regarding algorithmic decision-making, see Christopher Starke and others, 'Fairness Perceptions of Algorithmic Decision-Making: A Systematic Review of the Empirical Literature' (2022) 9 Big Data & Society 205395172211151 <http://journals.sagepub.com/doi/10.1177/20539517221115189> accessed 25 July 2025.
[4] Starke and others (n 3) 10.
[5] Maiju Tanninen, 'Contested Technology: Social Scientific Perspectives of Behaviour-Based Insurance' (2020) 7 Big Data & Society 205395172094253, 10 <http://journals.sagepub.com/doi/10.1177/2053951720942536>.
[6] Arthur Charpentier and Xavier Vamparys, 'Artificial Intelligence and Personalization of Insurance: Failure or Delayed Ignition?' (2025) 12 Big Data & Society 20539517241291817, 7–10 <https://journals.sagepub.com/doi/10.1177/20539517241291817>.
[7] Examples of recent attempts at regulating AI are the US bill of rights *Blueprint for an AI Bill of Rights. Making Automated Systems Work for the American People* (The White house | OSTP 2022) <https://www.whitehouse.gov/wp-content/uploads/2022/10/Blueprint-for-an-AI-Bill-of-Rights.pdf>. and the EU AI Act Regulation (EU) 2024/1689 of the European Parliament and of the Council of 13 June 2024 laying down harmonised rules on artificial intelligence and amending Regulations (EC) No 300/2008, (EU) No 167/2013, (EU) No 168/2013, (EU) 2018/858, (EU) 2018/1139 and (EU) 2019/2144 and Directives 2014/90/EU, (EU) 2016/797 and (EU) 2020/1828 (Artificial Intelligence Act) 2024.
[8] See the discussion section.
[9] Scholars in other sectors seem to see data-driven decision making as possibly disruptive. See Clarissa Valli Buttow, 'Data-Driven Policy Making and Its Impacts on Regulation: A Study of the OECD Vision in the Light of Data Critical Studies' (2025) 16 European Journal of Risk Regulation 114 <https://www.cambridge.org/core/product/identifier/S1867299X24000734/type/journal_article>. Insurers have decades (some maybe centuries) of experience with using large amounts of data. Some scholars see insurance as "an insightful analogon for the social situatedness and impact of machine learning systems" Christian Fröhlich and Robert C Williamson, 'Insights From Insurance for Fair Machine Learning', *The 2024 ACM Conference on Fairness, Accountability, and Transparency* (ACM 2024) <https://dl.acm.org/doi/10.1145/3630106.3658914>. See also the work by Barry and Charpentier: Laurence Barry and Arthur Charpentier, 'Melting Contestation: Insurance Fairness and Machine Learning' (2023) 25 Ethics and Information Technology 49 <https://link.springer.com/10.1007/s10676-023-09720-y>.
[10] See also the suggestions for further research at the end of the paper.

## 2 BACKGROUND: AI IN INSURANCE AND POSSIBLY UNFAIR DIFFERENTIATION

The core business of insurers is underwriting risks: estimating the expected claims cost of a consumer via a risk assessment.[11] Insurers always discriminate, in the neutral sense, between groups of people. Many insurers use a form of AI to underwrite risks or to analyze past data. AI can be described as "the science and engineering of making intelligent machines, especially intelligent computer programs".[12] For more than a century, insurers have gathered data directly from the consumers, such as data about a consumer's age, smoking, or drinking habits. Insurers used questionnaires to gather consumer characteristics such as the type of car they drive.[13]

Now, insurers could analyze more data gathered from many different sources. Insurers seem captivated by two trends (i) First, insurers could analyze more and new types of data to assess risks more precisely: data-intensive underwriting. (ii) Second, insurers could the behavior of individual consumers in real-time: behavior-based insurance. For example, some car insurers offer a discount if the consumer agrees to being tracked by the insurer and drives safely. AI makes both types of practices easier to implement for insurers.[14]

### 2.1 Data-intensive underwriting

The first trend is data-intensive underwriting: insurers use and analyze more data for underwriting. With machine learning, insurers can find new correlations in data with which the insurer could more accurately predict the expected claim costs of certain consumer groups. Insurers could analyze new datasets to find many new correlations in data.[15]

Data-intensive underwriting comes with potential discrimination-related risks. A legal risk of data-intensive underwriting is that insurers could, accidentally, discriminate indirectly in an illegal way (disparate impact). For example, if an insurer uses a newly found correlation to set premiums, the insurer could inadvertently cause harm to certain ethnic groups, or other groups with legally protected characteristics.[16] There are other possibly unfair, or at least controversial, effects of data-intensive underwriting. Our paper focuses on those effects.

First, consumers may be confronted with premiums based on characteristics that the consumer has little influence on. For example, in the Netherlands, some insurers charged different car insurance premiums based on the consumer's house number. A consumer paid more if their house number included a letter: for instance, 10A or 40B.[17] Consumers may feel that they cannot reasonably influence their house number. Moving to another house to lower one's car insurance premium is hardly a serious option. Consumers can influence other characteristics more easily, such as the type of car they buy.

Second, insurers could use certain consumer characteristics to set premiums, while the consumer does not see the logic of using the characteristic.[18] In other words, the insurer could use seemingly illogical or irrelevant characteristics to set the premium. AI systems are good at finding correlations in large datasets. When insurers underwrite using more data, they might find many correlations in datasets that look strange to consumers. In theory an insurer could find a correlation between on the one hand the chance that a consumer files a claim, and on the other hand whether a consumer has an address with an even number, is born in a certain month, or "people who spend more than 50% of their days on streets starting with the letter J".[19]

---

[11] The Geneva Association (Noordhoek), *Regulation of Artificial Intelligence in Insurance: Balancing Consumer Protection and Innovation* (The Geneva Assocation 2023) <https://www.genevaassociation.org/publication/public-policy-and-regulation/regulation-artificial-intelligence-insurance-balancing> accessed 25 July 2025.
[12] John McCarthy, 'What Is AI? Basic Questions' <http://jmc.stanford.edu/artificial-intelligence/what-is-ai/index.html> accessed 25 July 2025.
[13] Laurence Barry and Arthur Charpentier, 'Personalization as a Promise: Can Big Data Change the Practice of Insurance?' (2020) 7 Big Data & Society 205395172093514, 3–6 <http://journals.sagepub.com/doi/10.1177/2053951720935143>. Dan Bouk, 'How our days became numbered: Risk and the rise of the statistical individual', University of Chicago Press, 2019.
[14] See for more details on the two trends: Van Bekkum M, Zuiderveen Borgesius F en Heskes T, 'AI, Insurance, Discrimination and Unfair Differentiation: An Overview and Research Agenda' [2025] Law, Innovation and Technology <https://www.tandfonline.com/doi/full/10.1080/17579961.2025.2469348>.
[15] Autoriteit Financiële markten (AFM) and De Nederlandse Bank (DNB), *Artificial Intelligence in the Insurance Sector. An Exploratory Study* (AFM & DNB 2019) <https://www.afm.nl/~/profmedia/files/rapporten/2019/afm-dnb-verzekeringssector-ai-eng.pdf>.
[16] European Commission. Directorate General for Justice and Consumers. and others, *Indirect Discrimination under Directives 2000/43 and 2000/78.* (Publications Office 2022) <https://data.europa.eu/doi/10.2838/93469>. Aidan James McLoughney and others, '"Emerging Proxies" in Information-Rich Machine Learning: A Threat to Fairness?', *2023 IEEE International Symposium on Ethics in Engineering, Science, and Technology (ETHICS)* (IEEE 2023) <https://ieeexplore.ieee.org/document/10155045/> accessed 25 July 2025.
[17] Consumentenbond, 'Premies verzekeringen verschillen tot op huisnummer' ['Insurance premiums differ by house number'] (2015) <https://www.consumentenbond.nl/inboedelverzekering/verzekeringspremies-verschillen-tot-op-huisnummer> accessed 25 July 2025.
[18] Barry and Charpentier (n 9) 49.
[19] Cathy O'Neil, *Weapons of Math Destruction: How Big Data Increases Inequality and Threatens Democracy* (Penguin Books 2016) 138. See also Greta Krippner & Daniel Hirschman, 'The person of the category: The pricing of risk and the politics of classification in insurance and credit', *Theory and Society*, *51*(5), 685-727.

Third, insurers could introduce insurance products that are only intended for specific groups. Suppose, for example, that according to an insurer's analysis, people with a higher education submit fewer claims to their car insurer. The insurer could decide to offer car insurance only to a highly educated people. To illustrate, Promovendum, a Dutch insurer, markets its insurances as being only for the high-educated.[20] The insurer explains its reasoning: "Based on our years of experience and our claims statistics, we know that this target group has fewer claims. This enables us to offer our customers insurance products with low premiums and excellent terms and conditions".[21] Another insurer offers a car insurance only to medical doctors.[22] Such insurance practices can be controversial because the insurer targets specific groups of the population and thereby excludes the rest.

Fourth, insurance practices could reinforce inequalities that already exist in society, such as the difference between rich and poor. Such an effect could occur if an insurer (on purpose or by accident) charges higher prices to poorer people. In sum, data-intensive underwriting could have several negative effects.

## 2.2 Behavior-based insurance

A second trend in insurance is behavior-based insurance: insurers can adapt the premium to individual consumers in real-time, based on how the consumer behaves. For example, some car insurers offer a discount if the consumer agrees to being tracked by the insurer (with a device in the car) and drives safely.[23] A life insurer could give discounts to people who show, with a health tracker or smart watch, that they lead an active life.[24] Insurers can improve their behavior-based insurance with AI. For example, with a linear modelling approach, an insurer can consider characteristics such as weather and light conditions, mileage, time, and driving habits such as speeding to assess the consumer's driving behavior.[25]

The most important difference between data-intensive underwriting and behavior-based insurance is that with behavior-based insurance, the insurer bases the premium on the behavior of an *individual* consumer.[26] The insurer monitors, for instance, how the consumer drives, or how active a consumer is every day, measured by a health tracker or step counter. Another difference is that with behavior-based insurance, the insurer focuses more on preventing risks. For example, insurers could (at least in theory) prevent accidents by nudging risky drivers to drive safer.

Behavior-based insurance comes with partly similar risks as data-intensive underwriting. For example, some people may be excluded from behavior-based insurance. For bad drivers, behavior-based car insurance might be so expensive that, practically speaking, they are excluded from car insurance. In a hypothetical market where all car insurers offer behavior-based insurance, bad drivers may not have the possibility to insure themselves for an affordable price.[27]

Second, behavior-based insurance could reinforce financial inequality. Suppose for example, that on average, people with higher salaries have more time for sports or an active lifestyle. If richer people receive more discounts (based on their health trackers), they pay less. If richer people pay less, financial inequality is reinforced.

---

[20] The insurer does not directly refuse low-educated consumers requesting insurance. However, in advertisements and on the website, the insurer presents itself as being only for the higher-educated.
[21] Dutch text translated to English by the authors. Promovendum, 'Higher educated ['Hoger opgeleiden']' (*Promovendum*) <https://www.promovendum.nl/hoger-opgeleiden> accessed 25 July 2025.
[22] Axxellence, 'De Artsen-AutoVerzekering van Axxellence Nu Extra Voordelig!' ['Axxellence's Physician Car Insurance Now Extra Cheap!'] (2025) <https://www.axxellence.nl/auto.html> accessed 25 July 2025.
[23] Gert Meyers and Ine Van Hoyweghen, '"Happy Failures": Experimentation with Behaviour-Based Personalisation in Car Insurance' (2020) 7 Big Data & Society 205395172091465 <http://journals.sagepub.com/doi/10.1177/2053951720914650>.
[24] For a practical example, see The Vitality Group, 'The Vitality Difference', http://www.vitalitygroup.com/why-vitality/ accessed 23 July 2025. For a discussion of the case study, see Bednarz, Lewis & Sadowski, ''It's not personal, it's strictly business': Behavioural insurance and the impacts of non-personal data on individuals, groups and societies', *Computer Law & Security Review* 2025/56, s. 2. For a table with examples, see A Spender and others, 'Wearables and the Internet of Things: Considerations for the Life and Health Insurance Industry' (2019) 24 British Actuarial Journal e22 <https://www.cambridge.org/core/product/identifier/S1357321719000072/type/journal_article>, Table 4. See on behavior-based health insurance also also Gert Meyers, 'Behaviour-based Personalisation in Health Insurance: a Sociology of a not-yet Market' (KU Leuven 2018) <https://lirias.kuleuven.be/2087689&lang=en>.
[25] Dimitrios I Tselentis, George Yannis and Eleni I Vlahogianni, 'Innovative Motor Insurance Schemes: A Review of Current Practices and Emerging Challenges' (2017) 98 Accident Analysis & Prevention 139, 141 <https://linkinghub.elsevier.com/retrieve/pii/S0001457516303670>. For a laymen explanation of linear models in AI, see for example Katie Gross, 'Machine Learning and Linear Models: How They Work (In Plain English)' (2020) <https://blog.dataiku.com/top-machine-learning-algorithms-how-they-work-in-plain-english-1> accessed 25 July 2025.
[26] Insurers do not merely look at individuals when pricing behavior-based insurance; they also look at groups. See for a discussion: Liz Moor and Celia Lury, 'Price and the person: Markets, discrimination, and personhood', *Journal of Cultural Economy* 11.6 (2018): 501-513.
[27] Dutch Financial Authority (AFM), *The Personalisation of Prices and Conditions in the Insurance Sector. An Exploratory Study* (AFM 2021) <https://www.afm.nl/en/sector/actueel/2021/juni/aandachtspunten-gepersonaliseerde-beprijzing>. Tzameret H Rubin, Tor Helge Aas and Jackie Williams, 'Big Data and Data Ownership Rights: The Case of Car Insurance' (2023) 13 Journal of Information Technology Teaching Cases 82 <https://doi.org/10.1177/20438869221096859>.

In conclusion, both data-intensive underwriting and behavior-based insurance could have several possible negative effects. We do not claim that all these effects will occur; we merely highlight possible effects. We also note that insurers could combine both trends: insurers can aggregate individual behavioral data, and use the data for more data-intensive underwriting.[28] Insurers could then use that profit for more behavior-based insurance.[29] Nevertheless because of their partly different effects, we present the two trends separately.

## 3 METHOD

### 3.1 Survey design and sample

We conducted an online survey about data-intensive underwriting and behavior-based insurance among 1,000 Dutch citizens of 18 years and older, using a sample provided by panel service Panelclix in the Netherlands. In total, 1,102 respondents completed the questionnaire. We excluded respondents who completed the questionnaire in under 4 minutes ($n$=96), and respondents failing both attention checks ($n$=6).[30] One respondent requested to have their data removed after data collection; therefore, our final sample was $N$=999.

The final sample consisted of 51.6% female, 46.6% male, and 1.7% other/do not want to say. Mean age was 49.82 ($SD$=17.13). Of the sample 30.5% was highly educated, 51.7% received mid-level education, and 17.7% received primary or lower-level secondary education (Dutch population: 36.4, 37.0, and 26% respectively; CBS, 2024). Net monthly income was as follows: 6.9% earned less than €1,000, 35.6% between €1,000 and €2,500, 35.4% between €2,500 and €5,000, and 4,8% over €5,000 (net modal monthly income in the Netherlands in 2024 was €2.692; CPB, 2024). Hence, the sample consisted of a diverse demographic in terms of age, gender, education, and income.

Respondents filled out a questionnaire on "practices in the insurance sector". The median completion time was around 8 minutes. Respondents were compensated for their participation. The study complied with the criteria of Ethics Committee of the authors' institution (reference no. ECSW-LT-2025-1-10-82717). All data and materials are available at OSF.[31]

### 3.2 Procedure

Before respondents started the questionnaire, we asked them to give their informed consent. Subsequently, we asked respondents for their age. Next, respondents indicated their attitudes towards several short scenarios regarding data-intensive underwriting and behavior-based insurance, related to several insurance domains (e.g., car, life, or burglary insurance).

We included two attention checks in the questionnaire by means of instructed response items. The first check was an embedded item, instructing the respondents to "click fully agree here". A second check instructed respondents to click "green" in a list of colors.[32] Finally, respondents answered questions on socio-demographics. After completion, respondents were thanked and redirected to Panelclix collect their compensation. Respondents were paid 0,75 Euro for their participation.

### 3.3 Questionnaire

We presented respondents with various scenarios about different data-intensive insurance practices. Respondents read a short introduction, after which they were presented with several one-sentence examples of practices. For example, in one section ("block") of the survey we told respondents:

---

[28] Roel Henckaerts and Katrien Antonio, 'The Added Value of Dynamically Updating Motor Insurance Prices with Telematics Collected Driving Behavior Data' (2022) 105 Insurance: Mathematics and Economics 79 <https://linkinghub.elsevier.com/retrieve/pii/S0167668722000385>. Sobhan Moosavi and Rajiv Ramnath, 'Context-Aware Driver Risk Prediction with Telematics Data' (2023) 192 Accident Analysis & Prevention 107269 <https://linkinghub.elsevier.com/retrieve/pii/S0001457523003160>. In her book, O'Neil warns that "oceans of behavioral data will feed straight into artificial intelligence systems" O'Neil (n 19) 139.
[29] See, for example, recent work by Minty about the structural change of insurance and the chosen wording in recent debates. Duncan Minty, 'Insurance and the Importance of Words and Language' (*Ethics and Insurance*, 10 April 2025) <https://www.ethicsandinsurance.info/insurance-and-the-importance-of-words-and-language/> accessed 25 July 2025
[30] Marek Muszyński, 'Attention Checks and How to Use Them: Review and Practical Recommendations' (2023) 32 Ask: Research and Methods 3 <https://kb.osu.edu/handle/1811/103631> accessed 25 July 2025.
[31] <https://osf.io/x2fdy/?view_only=344f176daa404ec6929a21d576c4e159> accessed 25 July 2025.
[32] Muszyński (n 30).

*"This block is about how insurance companies determine the insurance premium (how much you pay each month). We want to know your opinion on this.*

*An insurance company (insurer) offers car insurance. The insurer calculates the monthly premium based on predictions of the number of claims for damages a consumer will make. The insurer makes these predictions based on information it has collected previously about the consumer, such as age, address, and type and year of the car. Below we present several examples of how an insurer can determine the premium. We ask for your opinion on each example.*

*Example 1: "The insurer determines the premium based on the postal code of the consumer."*

We wanted to know whether respondents found the practice that we described fair and acceptable. In other words, the two main outcome variables were Fairness and Acceptance. We measured *Fairness* by the statement: "I think this practice is fair". Respondents indicated their agreement on a 5-point Likert scale: (1) completely disagree; (2) disagree; (3) neutral; (4) agree; (5) completely agree. Likewise, for *Acceptance,* respondents indicated their agreement with the statement: "I think this practice is acceptable".

We also wanted to know whether respondents felt that a consumer could influence the premium, and whether they saw the logic behind the insurer using that characteristic to calculate the premium. Hence, additional variables were Perceived Influence and Perceived Logic. For *Influence* respondents indicated on a 5-point scale to what extent they agreed with the statement: "This way, the policyholder can influence the insurance premium". For *Logic*, respondents indicated to what extent they agreed with the statement "I find it logical that the insurer determines the premium like this". We asked the respondents the questions in Dutch. If we quote a question in this paper, we translate the question to English.

## 4 RESULTS

In this section, we present the results from the survey. Each subsection discusses a different research question of the paper. We present the results in text and figures. We include the full data in the tables in the Appendices.

### 4.1 Do people find (i) data-intensive underwriting and (ii) behavior-based insurance fair and acceptable?

The first question is: do respondents find data-intensive underwriting and behavior-based insurance fair and acceptable? A simple t-test compared the mean difference in fairness and acceptance between behavior-based insurance practices vs. practices based on data-intensive data types. Here, two data types classified as sources for behavior-based insurance, namely car tracker data ("car tracker") and data based on a step counter ("steps"), whereas the other data types classified as sources for data-intensive underwriting. The t-test indicated that behavior-based insurance practices overall score higher than data-intensive data types for insurance in terms of fairness (t (998) = 21.58, p < .001 (Mean Difference = .58)), and acceptance (t (998) = 18.26, p < .001 (Mean Difference = .51)).

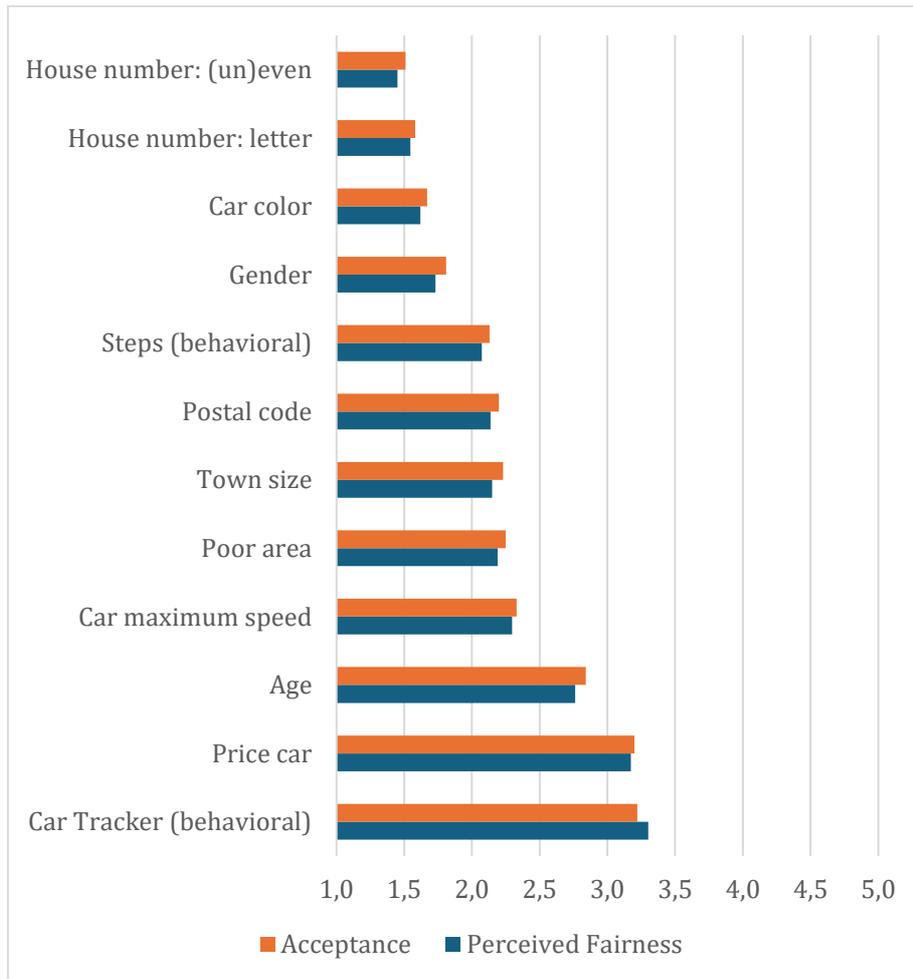

Figure 1. Public opinion of fairness and acceptance in insurance per data type

As Figure 1 shows, respondents found it only (relatively) fair (i.e., scoring above the scale midpoint of 3) if an insurer adapts the premium based on the consumer's driving behavior, which is an example of behavior-based insurance, and based on the maximum speed of the car, which is an example of data-intensive underwriting.[33] Respondents found it unfair and unacceptable if an insurer bases the premium on all the other types of data, including behavior-based insurance based on a step counter (i.e., scoring below the scale midpoint of 3). Hence, the significant difference between behavior-based insurance and data-intensive underwriting seems to be driven entirely by the higher fairness and acceptance values for the car tracker, and not by behavior-based insurance with a step counter. In conclusion, respondents find almost all modern insurance practices that we described unfair. See Appendix A for an overview of significance levels of deviations from the scale's neutral mid-point for each data type.

### 4.2 If people feel they have more influence on the premium, do they find an insurance practice fairer and more acceptable?

As noted, insurers could calculate premiums based on characteristics that consumers cannot reasonably influence (see section 2). We asked respondents whether they felt that they could influence the premium if the insurer calculated the premium based on the following characteristics: the consumer's gender, age, or postal code, the car's price, color, or maximum speed, the size of the city where the consumer lives, whether the consumer lived at an odd or even house number, and whether the consumer lived at a house number with a letter extension.

---

[33] See Section 3.3 (Questionnaire) for the description of the scenario. See also appendix D for life and burglary insurance.

We also asked respondents whether they find it fair and acceptable if an insurer calculates the premium based on such characteristics.

We asked two questions about life insurance (we include the introductions in Appendix D). In the first example, the insurer charges higher prices to poorer people. We asked respondents whether they felt that the consumer can influence the premium in such a case. The question is based on a real-life example of an insurer charging higher premiums in poor neighborhoods on purpose.[34]

We asked a second question about this life insurance example, but with an extra explanation: "In the Netherlands, people with a migration background are on average poorer than those without a migration background. This insurer's practice results in people with a migration background paying a higher premium on average than those without a migration background." (The example is based on a real-life insurance practice.[35]) Again, we asked respondents whether they thought that the consumer can influence the insurance premium.

We asked a similar question about burglary insurance.

We also asked respondents whether they felt that they could influence the insurance premium for two examples of behavior-based insurance: behavior-based car insurance (with a type of tracker in the car), and life insurance that gives a discount if the consumer walks more than 10.000 steps a day, as measured by a step counter.

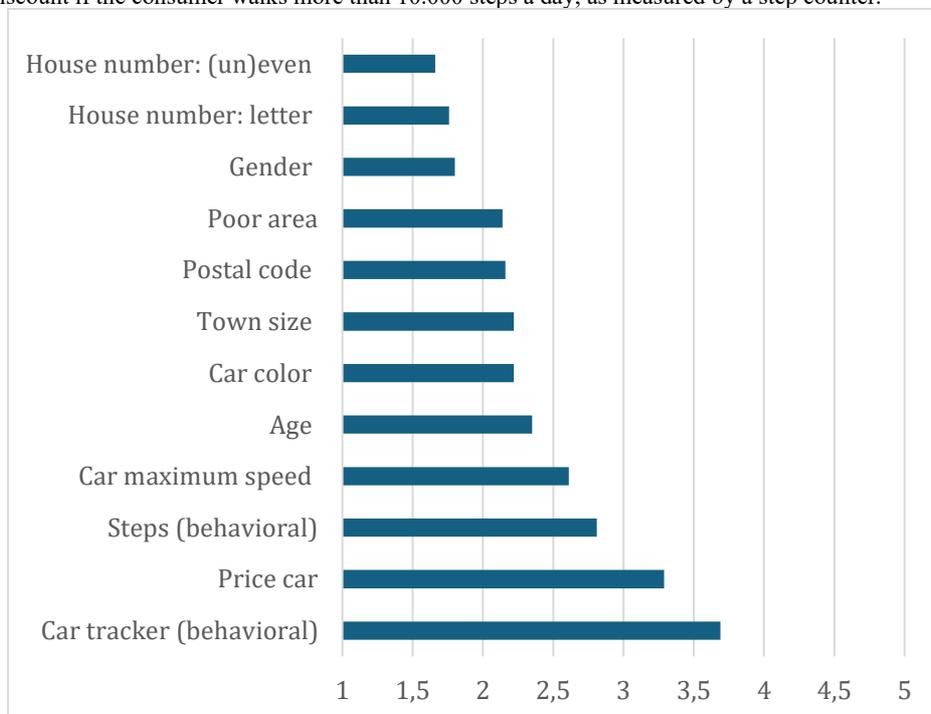

Figure 2. Influence perceptions by data type

Figure 2 shows the influence perceptions across all data types. For most data types, respondents feel that they can hardly influence the insurance premium. A one-sample t-test showed all values significantly deviated from the scale's midpoint (3; See Appendix B for significance levels for each data type). Most practices deviate negatively from this neutral point (where respondents find the practices neither highly nor little influenceable; Figure 2; See also Appendix B). This means that for most practices, respondents feel that the consumer cannot really influence the premium. However, respondents feel that consumers have some influence on the premium in the case of behavior-based insurance based on a car tracker and in the case of the price of the car they buy. Respondents do not feel that consumers have much influence on the premium in the case of behavior-based insurance with a step counter.

---

[34] College voor de Rechten van de Mens, *Advies aan Dazure B.V. over premiedifferentiatie op basis van postcode bij de Finvita overlijdensrisicoverzekering* (2014) <https://publicaties.mensenrechten.nl/publicatie/29c613ac-efab-4d2a-a26c-fe1cb2556407>.
[35] ibid.

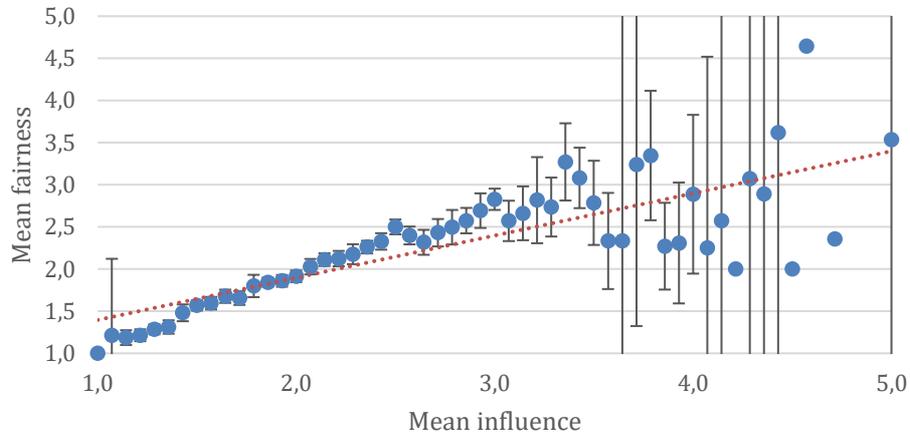

Figure 3. The relationship between perceived influence and fairness across all cases

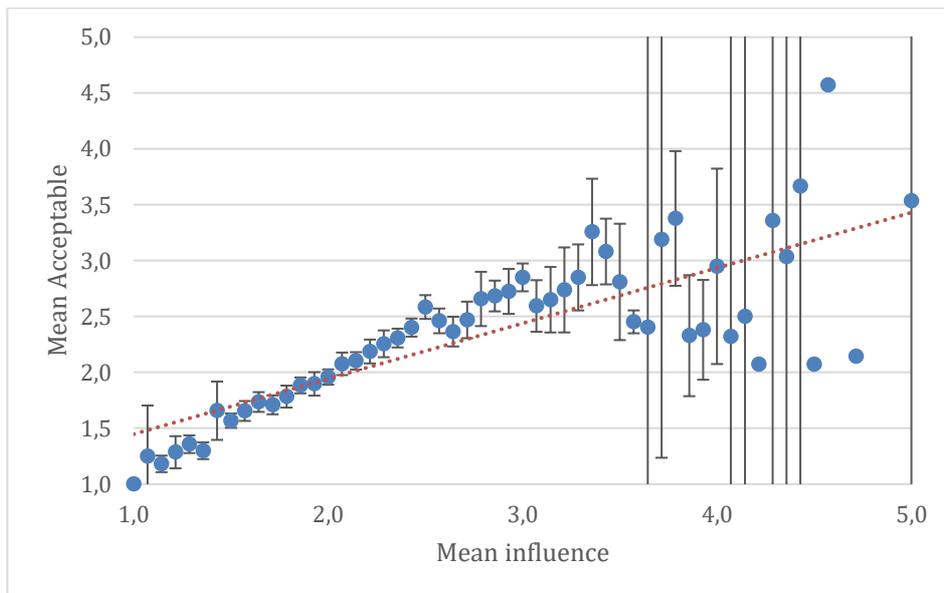

Figure 4. The relationship between perceived influence and acceptance across all cases36 below

We found a positive relationship between perceived influence and fairness and acceptance. Figures 3 and 4 give a graphic representation of the relationship between, on the one hand, perceived influence, and on the other hand perceived fairness (Figure 3) and acceptance (Figure 4). Regression analysis confirms the pattern that is shown in Figure 3, indicating that there is a positive relationship between perceived influence and the perceived fairness of the insurance practice across all cases ($R^2 = .54$, $F(1, 997) = 1147.52$, $\beta = .732$, $p < .001$). A similar conclusion can be drawn with regards to acceptance. Namely, as Figure 4 suggests, there is a positive relationship between perceived influence and perceived acceptance of the insurance practice ($R^2 = .53$, $F(1, 997) = 1126.68$, $\beta = .728$, $p < .001$).

In addition, Figures 3 and 4 suggest that if respondents feel that they have little influence on the insurance premium, then almost all respondents find the practice unfair (Figure 3) and unacceptable (Figure 4). However, if respondents find that they have more influence on the premium, then some find the practice fairer than others; hence, respondents are more divided. We can see this in the error bar in the figure, which is more volatile, or varied, for data types for which respondents feel that they have more influence on the premium.[36]

### 4.3 If people find an insurance practice more logical, do they find the practice fairer and more acceptable?

We asked respondents whether they saw the logic of certain insurance practices (See Section 2 about data-intensive underwriting). For various insurance practices, we asked respondents whether they found it logical that the insurer uses certain consumer characteristics to set the insurance premium. We also asked respondents whether they find it fair and acceptable if an insurer calculates the premium based on such practices.

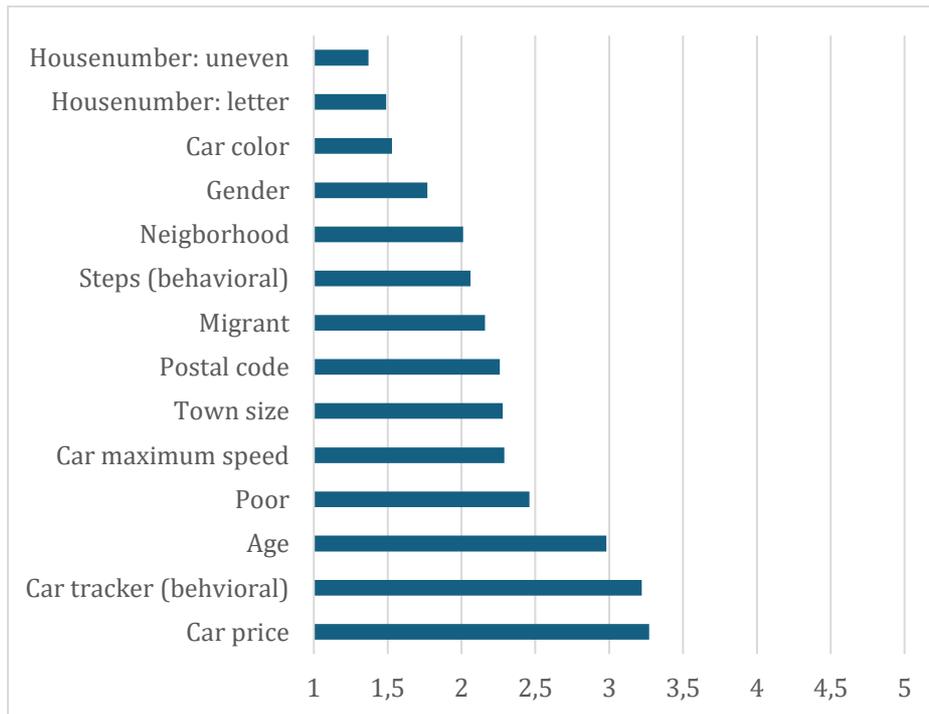

Figure 5. Perceived logic over all insurance practices

In Figure 5, we show the perceived logic for all insurance practices. Respondents considered a situation where the insurer uses the price of the insured car as the most logical. Respondents considered a situation where an insurer uses the type of house number to calculate the premium for car insurance the least logical (i.e., an uneven house number and, to a slightly lesser extent, a house number with a letter).[37] Compared to the scales' neutral midpoint of 3, respondents only considered situations in which car tracker data, or the price of a car, determined insurance fees as relatively logical (p's <.001). Respondents considered all other situations as relatively illogical (p's <.001), except for age, which the respondents considered neutral (p = .293; see Appendix B2 for all individual effects per insurance practice).

---

[36] Note. The scatterplot and regression line represent the relationship between perceived influence and acceptance of all cases of insurance practices (i.e., data-intensive underwriting, behavior-based practices and practices with indirect effects). The error bar is the vertical line attached to a dot in the figure. The error bar represents a 95% confidence interval. A 95% confidence interval means that there is only a 5% chance that the true value is NOT included within the span of the error bar.
[37] See Section 3.3 (Questionnaire) for the description of the scenario. See also appendix D for life and burglary insurance.

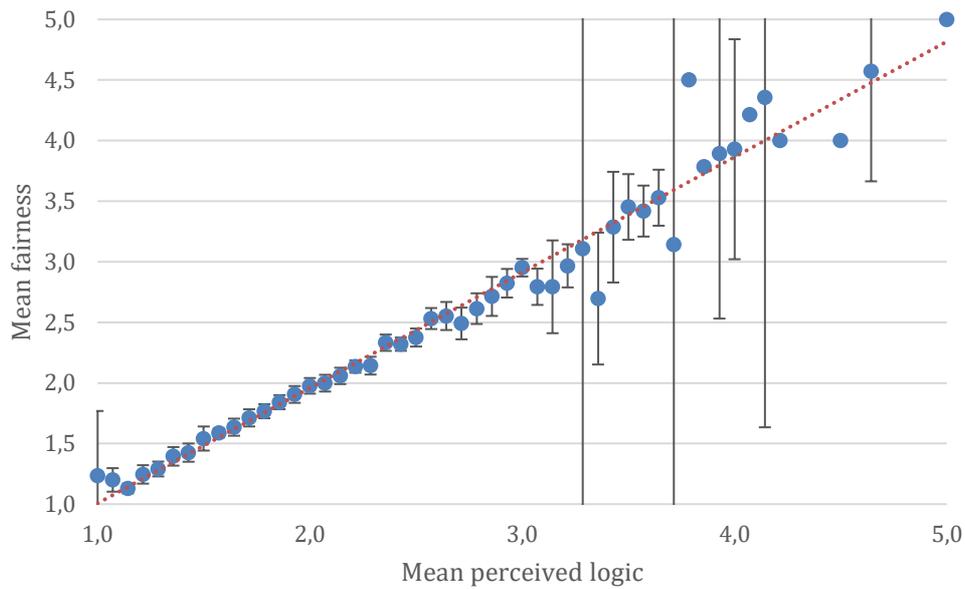

Figure 6. The relationship between perceived logic and fairness across all insurance practices

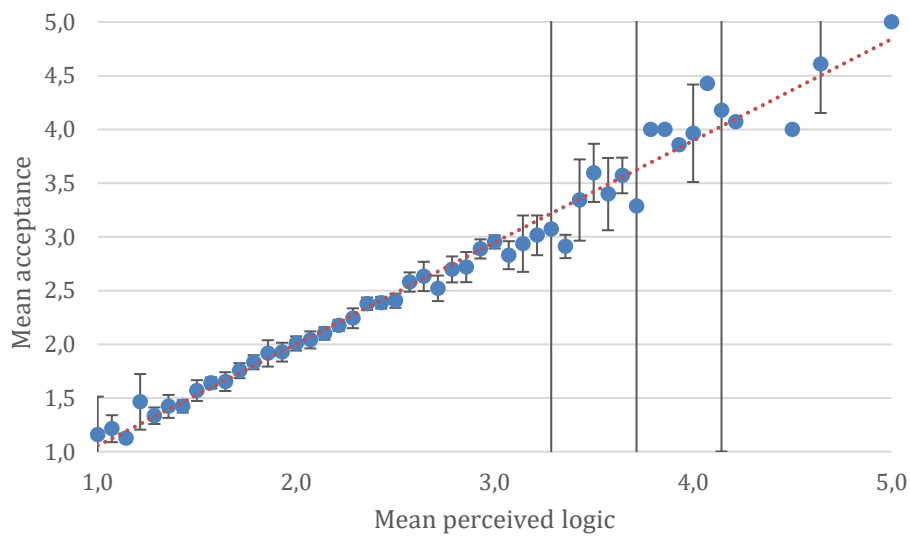

Figure 7. The relationship perceived logic and acceptance across all insurance practices

We found a strong positive relationship between perceived logic and fairness. The more logical respondents found a certain insurance practice, the fairer they thought the practice was ($R^2 = .84$, $F(1, 997) = 5142.06$, $\beta = .915$, $p < .001$). Similarly, we found a positive relationship between perceived logic and acceptance ($R^2 = .83$, $F(1, 997) = 4948.89$, $\beta = .912$, $p < .001$, see also Figure 6 and Figure 7). The error bars in the figures show that while respondents generally agree on the unfairness and unacceptability for examples that they find less logical, respondents disagree more often for more logical examples. In Section 4.3, we described a similar pattern for influence and fairness/acceptability. In sum, if respondents find an insurance practice more logical, they find the practice fairer and more acceptable.

## 4.4 If an insurer only accepts certain parts of the population, do people find such practices less fair and acceptable?

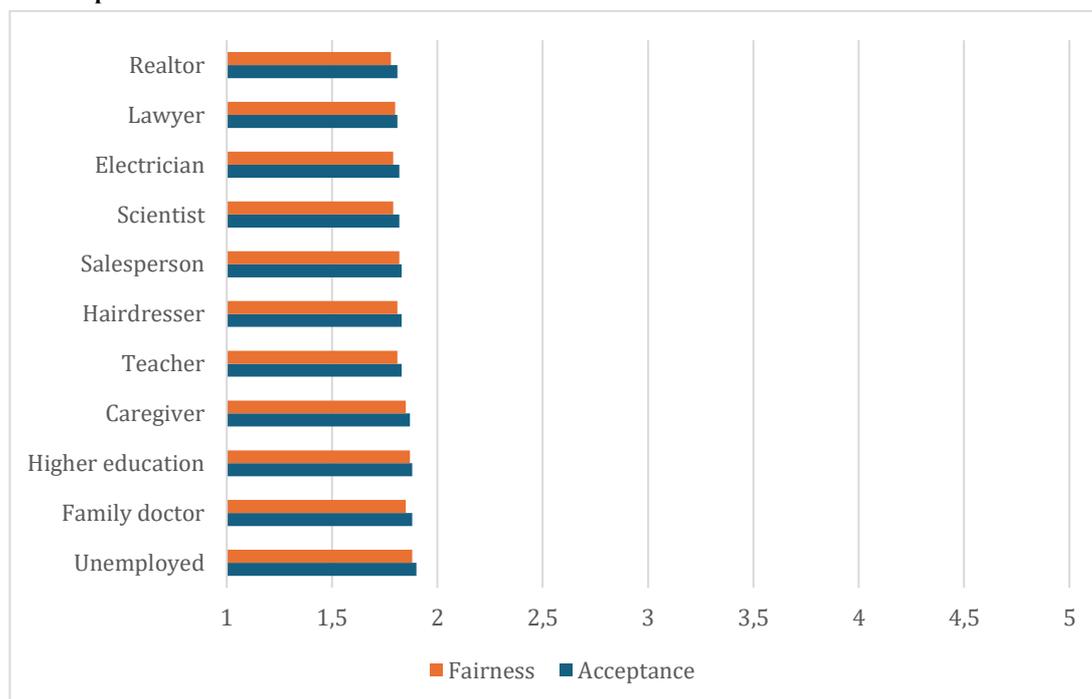

Figure 8. Respondents find target group insurance unfair and unacceptable

We asked whether respondents find it fair and acceptable if insurers only accept specific groups (see section 2). On average, respondents find all target group insurance completely unfair and unacceptable (all $p$'s <.001; see Figure 8; see also Appendix C for the individual significance levels of deviations from the scale's midpoint). It hardly matters for respondents who the target group is. And, when they assess whether they find a specific target group insurance fair and acceptable, it seems that respondents find it irrelevant whether the target group has a low income (caregiver, unemployed) or a high income (lawyer, family doctor).

## 4.5 If an insurance practice leads to higher prices for poorer consumers, do people find the practice less fair and less acceptable?

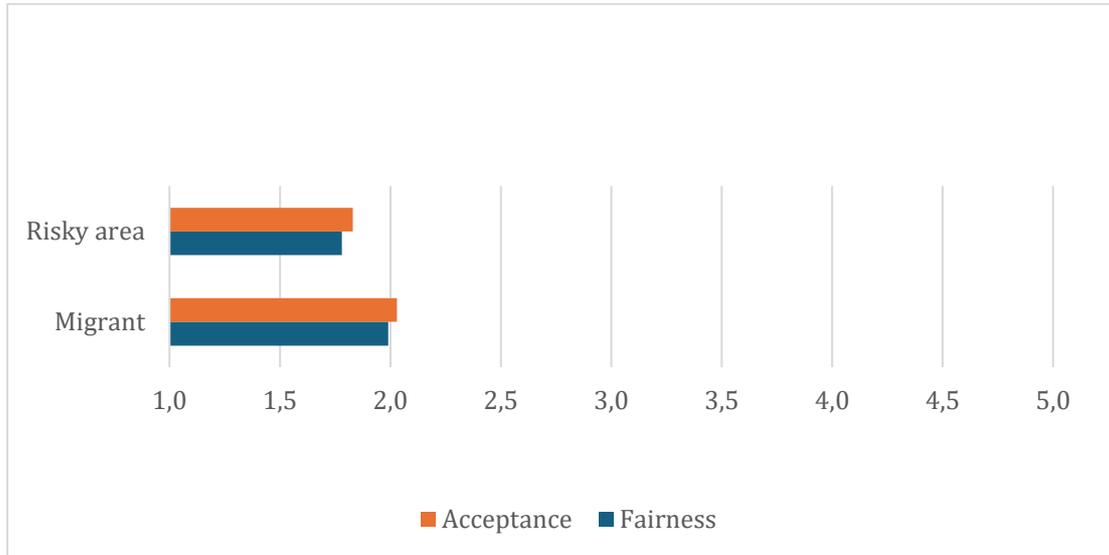

Figure 9. Public opinion of fairness and acceptance when poor population is indirect target group on the basis of…

Respondents find it unfair when insurance practices result in higher insurance premiums for poor people (Figure 9). We asked a question about an insurer who charges higher premiums in neighborhoods where more burglaries occur. We added: "This leads to poorer people paying higher premiums on average." (The full question is quoted in section 4.2). Hence, in this question, the fact that poorer people pay higher prices is an accidental effect of an insurance practice. The insurer did not charge higher premiums to poorer people on purpose, according to the question.

Regardless, respondents find the burglary insurance practice unfair and unacceptable (t (998) = -41.880, Mean Difference = -1.22, and unacceptable, t (998) = -38.27, Mean Difference = -1.17). See figure 9. So, at least in this example, respondents find it unfair and unacceptable if an insurance practice leads to higher prices for poorer consumers.

We also asked a question in which a life insurer charged a higher premium to poorer consumers on purpose. The reasoning of the insurer is that poorer people live shorter, and thus pay a premium for fewer years (see section 4.2. for the full question).

We asked a second question about life insurance that was largely the same, but included some extra information: "In the Netherlands, people with a migration background are on average poorer than people without a migration background. This insurer's practice results in people with a migration background paying a higher premium on average than those without a migration background." In a similar vein, respondents find it unfair if people with an immigrant background pay higher prices (t (998) = -29.19, p <.001, Mean Difference = -1.01), and unacceptable (t (998) = -27.78, Mean Difference = -.97). Figure 9 shows the results. All in all, respondents find it unfair and unacceptable if poorer consumers pay more for insurance.

## 5 DISCUSSION

In this section, we reflect on the main findings from the survey. First, broadly speaking, if respondents find a practice unfair, they also find the practice unacceptable. For easy of reading, we only speak about "fair" and "unfair" below, rather than about "fair and acceptable" and "unfair and unacceptable". Second, the survey respondents find almost all modern insurance practices that we described very unfair. Insurers who want to apply data-intensive underwriting or behavior-based insurance should be aware that many people dislike those practices. Behavior-based car insurance seems the least unfair of the practices: respondents found behavior-based car insurance slightly fairer than neutral. This gives some preliminary evidence for a standard theory in insurance fairness: some say that behavior-based insurance is more "actuarially fair" than traditional insurance.[38] A more expansive survey or experiment could confirm whether the same results hold.

Third, respondents find insurance practices fairer if they feel that they can influence the premium. This view of respondents is understandable. We can imagine that it feels unfair to be punished with a higher premium for your car insurance, just because you live in an apartment with a letter in the house number. Meanwhile, insurers sometimes set premiums based on such data (see section 2). In the future, insurers may use more data on which a consumer has little influence on to set premiums. By underwriting more data-intensively, insures could find more correlations that help them to predict claim costs. After all, machine learning is good at finding correlations in large datasets. It may be bad news for insurers that many consumers dislike it if insurers calculate premiums based on characteristics that the consumer cannot influence.

We asked about one other type of behavior-based insurance: life insurance with discounts based on a step counter. Respondents find such step counter insurance unfair. What could explain this difference between respondents' fairness perceptions about behavior-based car insurance and behavior-based insurance based on a step counter? We provide some preliminary thoughts. First, perhaps people feel that they cannot help it if they make few steps per day. Perhaps their legs hurt, or perhaps physical activity does not fit in their busy schedules. People might think that, by comparison, they have more influence on their driving style. Second, perhaps some people do physical activity (e.g. strength training) that a step counter does not register well. Such people might see a step counter as a bad way to register physical activity. Third, we asked about a discount if you register more than 10.000 steps. Perhaps people's reaction of respondents would have been different if we asked them about a different number of steps. More research on such questions would be welcome.

Fourth, respondents dislike it if insurers use seemingly illogical consumer characteristics to set premiums. A controversial characteristic that respondents found logical is Age. Perhaps in the scenario we presented, age was deemed fairer. Other surveys could ask about other situations. Respondents consider behavior-based the most logical, but this does not seem true for the step counter. This could be because of the scenario we presented (10.000 steps). Respondents find it unfair if an insurer calculates the premium based on a consumer characteristic when the respondent does not see the logic, or relevance, of using that characteristic. The results raise questions about whether insurers can underwrite data-intensively in a way that consumers find fair. Insurers may find it difficult to convince consumers of the logic or relevance of the data that they use for underwriting, even if in the eyes of the insurer, the data are relevant. Suppose that an insurer only wants to use consumer characteristics to set premiums if it can explain the logic or relevance of using that characteristic. Such a fairness policy might mean that the insurer cannot use certain consumer characteristics, even though an AI system has found a correlation between those characteristics and filing claims.

Fifth, respondents dislike target group insurance. Respondents find all types of insurance that are only available for certain groups, such as a car insurance specifically for family doctors, unfair. Meanwhile, some insurers do offer such target group insurance. Because in the Netherlands, target group insurance for doctors is long established, it is remarkable that respondents seem to consider all target group insurances equally unfair. Perhaps people feel discriminated if they are not allowed to buy certain insurance products. Perhaps people's opinions change if an insurer gives arguments for a certain type of target group insurance. Sixth, respondents find it unfair if poor people pay higher insurance prices. Maybe respondents feel that such practices discriminate against poor people.

An insurer could try to make practices fairer by using a technical "fairness metric" that helps to prevent some unfair practices. A fairness metric is, roughly speaking, a score that expresses how fair a system is in terms of, for example, the number of false positives and false negatives).[39] However, for the examples we used in our survey, many fairness metrics do not seem suitable. Many metrics, such as Equality of Odds, focus on the outcome rather than whether a characteristic seems logical or is influenceable for the consumer.[40] Such metrics are useful, for example, if an insurer wishes to prevent the poor paying more. Perhaps metrics based on causality or control could provide some insight into the logic behind certain characteristics and whether consumers can influence the characteristic. But even then, it may be difficult for an insurer to foresee what characteristics and practices consumers see as logical.[41] Similar to recent fairness research, we think that bringing down fairness to a single metric is difficult.[42] Explainability may help insurers understand the correlations they come across while applying machine learning to new datasets. But even if insurers can explain a decision to the consumer, that explanation does not necessarily make the decision itself fairer.[43] For example, even if an insurer explains why it finds a house number relevant for car insurance, the consumer may disagree with the explanation. Most consumers are not statisticians.[44]

---

[38] Van Bekkum M, Zuiderveen Borgesius F and Heskes T, 'AI, Insurance, Discrimination and Unfair Differentiation: An Overview and Research Agenda' [2025] Law, Innovation and Technology https://www.tandfonline.com/doi/full/10.1080/17579961.2025.2469348, s 5.3. See also Gert Meyers and Ine Van Hoyweghen, 'Enacting Actuarial Fairness in Insurance: From Fair Discrimination to Behaviour-Based Fairness' (2018) 27 Science as Culture 413 <https://www.tandfonline.com/doi/full/10.1080/09505431.2017.1398223>.
[39] Solon Barocas, Moritz Hardt and Arvind Narayanan, *Fairness and Machine Learning* (MIT Press 2023) 54 <https://fairmlbook.org/>.
[40] Solon Barocas, Moritz Hardt and Arvind Narayanan, *Fairness and Machine Learning* (MIT Press 2023) <https://fairmlbook.org/>.

It seems that the law in Europe can hardly protect people against the unfairness types highlighted in this paper. For instance, most insurance practices described in this paper do not violate non-discrimination law. Most non-discrimination statutes only protect people against discrimination based on certain protected characteristics, such as age, disability, gender, religion, ethnicity, or sexual orientation. Through the concept of indirect discrimination (disparate impact) some insurance practices might be illegal, if they disproportionally harm people with a certain ethnicity or other protected characteristic.[45] Most rules in the AI Act only apply to health and life insurance, not to other types of insurance. And even for health and life insurance, the AI Act does not ban or clearly regulate the practices described in this paper.[46] EU law specific for insurance does not focus on the unfairness types in this paper either.[47] The best source for some, albeit limited, legal protection is probably the General Data Protection Regulation.[48] The GDPR may limit the amount of data that insurers can gather and analyze, and the GDPR's transparency requirements could help to make some insurance practices less of a black box.[49] The GDPR's rules on certain fully automated decisions with legal or similar effects may also offer some protection.[50] But the GDPR does not outright ban data-intensive underwriting or behavior-based insurance. All in all, it appears that current law can offer, at most, some limited protection against the perceived unfairness discussed in this paper. An in-depth analysis of possible legal protection falls outside the scope of this paper.

We do not think that policymakers should ban every practice that people dislike or find unfair. To illustrate: many people dislike paying taxes. But the state needs tax income to run the country. Policymakers should consider many aspects of a problem. Similarly, we do not think that a practice is always morally fine, if a majority agrees with the practice. In our opinion, policymakers should give some attention to what people find fair and acceptable according to surveys, as long as policymakers scrutinize the survey in question, e.g. for data quality.[51]

Finally, this paper did not show that, according to the respondents, certain insurance practices must be prohibited by a lawmaker or withdrawn by an insurer. What people consider unacceptable may not necessarily overlap with their buying decisions - a "fairness paradox". People's stated preferences (what people say in surveys) may differ from their observed behavior (how people act). For example, somebody might say that it is unacceptable if a life insurer offers discounts to people who show an active lifestyle with a step counter, but that same person might buy such an insurance product because the option is more affordable. Related discussions exist about a "privacy paradox": stated privacy preferences versus privacy behavior.[52] Even if a fairness paradox exists, our paper can still inform the debate about the fairness and acceptance of data-intensive underwriting and behavior-based insurance.

---

[41] Van Bekkum M, Zuiderveen Borgesius F and Heskes T, 'AI, Insurance, Discrimination and Unfair Differentiation: An Overview and Research Agenda' [2025] Law, Innovation and Technology https://www.tandfonline.com/doi/full/10.1080/17579961.2025.2469348, s 4.2.2. See also Fröhlich and Williamson (n 9) 411.
[42] See e.g. Tim De Jonge and Djoerd Hiemstra, 'UNFair: Search Engine Manipulation, Undetectable by Amortized Inequity', *2023 ACM Conference on Fairness, Accountability, and Transparency* (ACM 2023) 837 <https://dl.acm.org/doi/10.1145/3593013.3594046> accessed 25 July 2025.
[43] For an overview of the similarities and differences between fairness and explainability, see Luca Deck and others, 'A Critical Survey on Fairness Benefits of Explainable AI', *The 2024 ACM Conference on Fairness, Accountability, and Transparency* (ACM 2024) <https://dl.acm.org/doi/10.1145/3630106.3658990> accessed 25 July 2025.
[44] See also Van Bekkum M, Zuiderveen Borgesius F and Heskes T, 'AI, Insurance, Discrimination and Unfair Differentiation: An Overview and Research Agenda' [2025] Law, Innovation and Technology https://www.tandfonline.com/doi/full/10.1080/17579961.2025.2469348, p. 16.
[45] Van Bekkum M, Zuiderveen Borgesius F and Heskes T, 'AI, Insurance, Discrimination and Unfair Differentiation: An Overview and Research Agenda' [2025] Law, Innovation and Technology https://www.tandfonline.com/doi/full/10.1080/17579961.2025.2469348, s 4.1 & 5.1.
[46] Regulation (EU) 2024/1689 of the European Parliament and of the Council of 13 June 2024 laying down harmonised rules on artificial intelligence and amending Regulations (EC) No 300/2008, (EU) No 167/2013, (EU) No 168/2013, (EU) 2018/858, (EU) 2018/1139 and (EU) 2019/2144 and Directives 2014/90/EU, (EU) 2016/797 and (EU) 2020/1828 (Artificial Intelligence Act).
[47] Directive (EU) 2016/97 of the European Parliament and of the Council of 20 January 2016 on insurance distribution (recast) (2016/97).
[48] Regulation (EU) 2016/179 of the European Parliament and the Council on the protection of natural persons with regard to the processing of personal data and on the free movement of such data, and repealing Directive 95/46/EC (General Data Protection Regulation).
[49] Article 12-14, General Data Protection Regulation.
[50] Article 22, General Data Protection Regulation.
[51] Hawal Shamon and Carl Clemens Berning, 'Attention Check Items and Instructions in Online Surveys: Boon or Bane for Data Quality?' [2020] Survey Research Methods 55 <https://ojs.ub.uni-konstanz.de/srm/article/view/7374>. Scott Clifford and Jennifer Jerit, 'Cheating on Political Knowledge Questions in Online Surveys: An Assessment of the Problem and Solutions' (2016) 80 Public Opinion Quarterly 858 <https://academic.oup.com/poq/article-lookup/doi/10.1093/poq/nfw030>.

## 6  LIMITATIONS AND SUGGESTIONS FOR FURTHER RESEARCH

In this section we highlight limitations of this research and give suggestions for further research. This survey research was a first attempt to learn more about people's opinions about two modern trends in insurance. The sample of 999 respondents gives a reasonable cross-section of Dutch society, but the sample is not fully representative.

There are exciting opportunities for further research, in many disciplines. Social scientists could do more empirical research. Survey research in more countries could highlight the differences between countries. And surveys could contain more, and more in-depth, questions about issues such as behavior-based insurance, health insurance, and insurance practices that reinforce financial inequality. Focus groups or interviews could provide more insight into why people find certain practices unfair. Further research could investigate whether there is there a "fairness paradox" in relation to data-intensive underwriting and behavior-based insurance, i.e. the relationship between people's stated fairness preferences and their buying decisions.

Computer scientists could analyze whether fairness metrics, as proposed by many papers of the ACM FAccT Conference for example, could help to protect people against the types of (perceived) unfairness as highlighted in this paper.[53]

There are many open normative and ethical questions. For instance, from a normative perspective, when is it acceptable for insurance practices to increase financial inequality? Is it fair for people to bear the cost of characteristics or behaviors that they have no or little control over or find illogical?[54] Could a comprehensive normative framework be created for insurance?[55] Is behavior-based insurance fairer than traditional insurance or data-intensive underwriting? Should insurers communicate the shift to more behavior-based insurance with language implying solidarity or with language implying rationality?[56]

Many legal questions deserve more research too. Perhaps an in-depth analysis of current law could show that the GDPR or other law offer some protection against the perceived unfairness discussed in this paper. And maybe additional regulation should be considered. Should policymakers do more to mitigate the risk that insurance practices reinforce social inequality? There is some precedent: in many situations, the law aims to protect the weaker party, such as in the field of labor law and consumer protection law. Are sector-specific rules needed for the use of AI in the insurance sector? Should the law do more to ensure that consumers are not excluded from certain insurance products, or should contract freedom stay the same? Normative and ethical research could discuss whether, and if so when, it is appropriate for the law to intervene.

## 7  CONCLUDING REMARKS

With a survey in the Netherlands, we explored people's opinions about two modern trends in insurance: (i) data-intensive underwriting and (ii) behavior-based insurance. The main results include the following. First, if survey respondents find an insurance practice unfair, they also find the practice unacceptable. Second, respondents find almost all modern insurance practices that we described unfair. Third, respondents find practices fairer if they can influence the premium. For example, respondents find behavior-based car insurance based on a car tracker relatively fair. Fourth, respondents find it unfair if an insurer calculates the premium based on a consumer characteristic, while the respondent does not see the logic of using that characteristic. Fifth, respondents find all types of insurance that are only available for certain groups, such as a car insurance specifically for family doctors, unfair. Sixth, respondents find it unfair if an insurance practice leads to higher prices for poorer people. Overall, respondents worry about these practices, finding many of them unfair. Many open questions remain, for many disciplines: we suggest further research for social scientists, legal scholars, ethicists, and computer scientists.

---

[52] Spyros Kokolakis, 'Privacy Attitudes and Privacy Behaviour: A Review of Current Research on the Privacy Paradox Phenomenon' (2017) 64 Computers & Security 122 <https://linkinghub.elsevier.com/retrieve/pii/S0167404815001017>. Daniel J Solove, 'The Myth of the Privacy Paradox' [2020] SSRN Electronic Journal <https://www.ssrn.com/abstract=3536265>. Lina Dencik and Jonathan Cable, 'The Advent of Surveillance Realism: Public Opinion and Activist Responses to the Snowden Leaks', *International Journal of Communication* 11 (2017): 763-781.
[53] One paper exploring such a topic is Fröhlich and Williamson (n 9).
[54] While some papers discuss this question in general, in the past, insurers rejected 'the imposition of a requirement that so-called 'risk-factors' must have a demonstrable causal relationship to variations in the cost of claims [.] largely because it is based on data mining' Oscar H Gandy, *Coming to Terms with Chance: Engaging Rational Discrimination and Cumulative Disadvantage* (Routledge 2016) 117.
[55] Avraham made a first contribution toward such a framework Ronen Avraham, Kyle D Logue and Daniel Schwarcz, 'Understanding Insurance Antidiscrimination Laws' [2014] 87 S. CAL. L. REV. 195 <https://scholarship.law.umn.edu/faculty_articles/576>.
[56] Minty (n 29).

**APPENDIX A**

Table A1. Fairness by data type

|  | t | df | Significance One-Sided p | Significance Two-Sided p | Mean Difference | 95% Confidence Interval of the Difference Lower | 95% Confidence Interval of the Difference Upper |
|---|---|---|---|---|---|---|---|
| Postal code | -25.042 | 998 | <.001 | <.001 | -.862 | -.93 | -.79 |
| Uneven | -59.407 | 998 | <.001 | <.001 | -1.551 | -1.60 | -1.50 |
| Letter | -51.713 | 998 | <.001 | <.001 | -1.453 | -1.51 | -1.40 |
| Price | 4.747 | 998 | <.001 | <.001 | .173 | .10 | .24 |
| Speed | -19.872 | 998 | <.001 | <.001 | -.703 | -.77 | -.63 |
| Color | -47.656 | 998 | <.001 | <.001 | -1.381 | -1.44 | -1.32 |
| Location | -25.630 | 998 | <.001 | <.001 | -.851 | -.92 | -.79 |
| Gender | -42.661 | 998 | <.001 | <.001 | -1.269 | -1.33 | -1.21 |
| Age | -6.999 | 998 | <.001 | <.001 | -.238 | -.31 | -.17 |
| Tracker | 8.408 | 998 | <.001 | <.001 | .302 | .23 | .37 |
| Steps | -27.432 | 998 | <.001 | <.001 | -.928 | -.99 | -.86 |
| Neighborhood | -41.880 | 998 | <.001 | <.001 | -1.219 | -1.28 | -1.16 |
| Migrant | -29.187 | 998 | <.001 | <.001 | -1.010 | -1.08 | -.94 |
| Poor | -23.398 | 998 | <.001 | <.001 | -.811 | -.88 | -.74 |

*Note. Negative t-values represent a negative deviation from the scale midpoint, whereas positive t-values represent positive deviations from the scale midpoint. Any mean differences are significant at the p<.001 level.*

Table A2. Acceptance by data-type

|  | t | df | Significance One-Sided p | Significance Two-Sided p | Mean Difference | 95% Confidence Interval of the Difference Lower | 95% Confidence Interval of the Difference Upper |
|---|---|---|---|---|---|---|---|
| Postal code | -22.345 | 998 | <.001 | <.001 | -.797 | -.87 | -.73 |
| Uneven | -51.931 | 998 | <.001 | <.001 | -1.494 | -1.55 | -1.44 |
| Letter | -48.404 | 998 | <.001 | <.001 | -1.417 | -1.47 | -1.36 |
| Price | 5.493 | 998 | <.001 | <.001 | .202 | .13 | .27 |
| Speed | -18.503 | 998 | <.001 | <.001 | -.674 | -.75 | -.60 |
| Color | -43.070 | 998 | <.001 | <.001 | -1.328 | -1.39 | -1.27 |
| Location | -21.368 | 998 | <.001 | <.001 | -.750 | -.82 | -.68 |
| Gender | -35.978 | 998 | <.001 | <.001 | -1.189 | -1.25 | -1.12 |
| Age | -4.639 | 998 | <.001 | <.001 | -.158 | -.23 | -.09 |
| Tracker | 6.141 | 998 | <.001 | <.001 | .223 | .15 | .29 |
| Steps | -25.161 | 998 | <.001 | <.001 | -.875 | -.94 | -.81 |
| Neighborhood | -38.270 | 998 | <.001 | <.001 | -1.171 | -1.23 | -1.11 |
| Migrant | -27.784 | 998 | <.001 | <.001 | -.970 | -1.04 | -.90 |
| Poor | -21.787 | 998 | <.001 | <.001 | -.768 | -.84 | -.70 |

*Note. Negative t-values represent a negative deviation from the scale midpoint, whereas positive t-values represent positive deviations from the scale midpoint. Any mean differences are significant at the p<.001 level.*

**APPENDIX B**

Table B1. Influence by data-type

|  | t | df | Significance | | Mean Difference | 95% Confidence Interval of the Difference | |
|---|---|---|---|---|---|---|---|
|  |  |  | One-Sided p | Two-Sided p |  | Lower | Upper |
| Postal code | -23.985 | 998 | <.001 | <.001 | -.840 | -.91 | -.77 |
| House number: uneven | -41.272 | 998 | <.001 | <.001 | -1.339 | -1.40 | -1.28 |
| House number: letter | -37.216 | 998 | <.001 | <.001 | -1.244 | -1.31 | -1.18 |
| Price car | 8.320 | 998 | <.001 | <.001 | .293 | .22 | .36 |
| Car maximum speed | -9.930 | 998 | <.001 | <.001 | -.394 | -.47 | -.32 |
| Car color | -18.669 | 998 | <.001 | <.001 | -.782 | -.86 | -.70 |
| Town size | -22.520 | 998 | <.001 | <.001 | -.785 | -.85 | -.72 |
| Gender | -35.177 | 998 | <.001 | <.001 | -1.204 | -1.27 | -1.14 |
| Age | -17.354 | 998 | <.001 | <.001 | -.647 | -.72 | -.57 |
| Car tracker (behavioral) | 19.994 | 998 | <.001 | <.001 | .693 | .62 | .76 |
| Steps (behavioral) | -4.478 | 998 | <.001 | <.001 | -.189 | -.27 | -.11 |
| Poor area | -24.879 | 998 | <.001 | <.001 | -.861 | -.93 | -.79 |

*Note. Negative t-values represent a negative deviation from the scale midpoint, whereas positive t-values represent positive deviations from the scale midpoint. Any mean differences are significant at the p<.001 level.*

Table B2. Perceived logic by insurance practice

|  | t | df | Significance One-Sided p | Significance Two-Sided p | Mean Difference | 95% Confidence Interval of the Difference Lower | 95% Confidence Interval of the Difference Upper |
|---|---|---|---|---|---|---|---|
| Postal code | -20.011 | 998 | <.001 | <.001 | -.739 | -.81 | -.67 |
| House number: even | -69.493 | 998 | <.001 | <.001 | -1.635 | -1.68 | -1.59 |
| House number: letter | -57.445 | 998 | <.001 | <.001 | -1.507 | -1.56 | -1.46 |
| Price car | 7.240 | 998 | <.001 | <.001 | .272 | .20 | .35 |
| Car maximum speed | -19.239 | 998 | <.001 | <.001 | -.714 | -.79 | -.64 |
| Car color | -54.640 | 998 | <.001 | <.001 | -1.473 | -1.53 | -1.42 |
| Location | -20.424 | 998 | <.001 | <.001 | -.725 | -.79 | -.66 |
| Gender | -38.598 | 998 | <.001 | <.001 | -1.232 | -1.29 | -1.17 |
| Age | -.544 | 998 | .293 | .587 | -.019 | -.09 | .05 |
| Car tracker (behavioral) | 6.056 | 998 | <.001 | <.001 | .222 | .15 | .29 |
| Steps (behavioral) | -27.671 | 998 | <.001 | <.001 | -.940 | -1.01 | -.87 |
| Neigborhood | -28.901 | 998 | <.001 | <.001 | -.993 | -1.06 | -.93 |
| Migrant | -22.874 | 998 | <.001 | <.001 | -.837 | -.91 | -.77 |
| Poor area | -14.108 | 998 | <.001 | <.001 | -.544 | -.62 | -.47 |

*Note. Negative t-values represent a negative deviation from the scale midpoint (except Age, which is not significant from the scale's midpoint), whereas positive t-values represent positive deviations from the scale midpoint. Any mean differences are significant at the p<.001 level.*

21**APPENDIX C**

Table C1. Fairness by target group

| | t | df | Significance One-Sided p | Two-Sided p | Mean Difference | 95% Confidence Interval of the Difference Lower | Upper |
|---|---|---|---|---|---|---|---|
| Family doctor | -35.159 | 998 | <.001 | <.001 | -1.147 | -1.21 | -1.08 |
| Lawyer | -37.321 | 998 | <.001 | <.001 | -1.199 | -1.26 | -1.14 |
| Scientist | -38.081 | 998 | <.001 | <.001 | -1.209 | -1.27 | -1.15 |
| Teacher | -36.478 | 998 | <.001 | <.001 | -1.187 | -1.25 | -1.12 |
| Realtor | -38.571 | 998 | <.001 | <.001 | -1.216 | -1.28 | -1.15 |
| Electrician | -38.210 | 998 | <.001 | <.001 | -1.206 | -1.27 | -1.14 |
| Caregiver | -34.704 | 998 | <.001 | <.001 | -1.147 | -1.21 | -1.08 |
| Salesperson | -36.369 | 998 | <.001 | <.001 | -1.178 | -1.24 | -1.11 |
| Hairdresser | -37.366 | 998 | <.001 | <.001 | -1.194 | -1.26 | -1.13 |
| Unemployed | -33.479 | 998 | <.001 | <.001 | -1.118 | -1.18 | -1.05 |
| Higher education | -33.849 | 998 | <.001 | <.001 | -1.133 | -1.20 | -1.07 |

*Note. Negative t-values represent a negative deviation from the scale midpoint, whereas positive t-values represent positive deviations from the scale midpoint. Any mean differences are significant at the p<.001 level.*

Table C2. Acceptance by target group

| | t | df | Significance One-Sided p | Two-Sided p | Mean Difference | 95% Confidence Interval of the Difference Lower | Upper |
|---|---|---|---|---|---|---|---|
| Family doctor | -33.177 | 998 | <.001 | <.001 | -1.124 | -1.19 | -1.06 |
| Lawyer | -36.683 | 998 | <.001 | <.001 | -1.192 | -1.26 | -1.13 |
| Scientist | -36.179 | 998 | <.001 | <.001 | -1.184 | -1.25 | -1.12 |
| Teacher | -35.322 | 998 | <.001 | <.001 | -1.174 | -1.24 | -1.11 |
| Realtor | -35.971 | 998 | <.001 | <.001 | -1.187 | -1.25 | -1.12 |
| Electrician | -35.563 | 998 | <.001 | <.001 | -1.178 | -1.24 | -1.11 |
| Caregiver | -33.135 | 998 | <.001 | <.001 | -1.131 | -1.20 | -1.06 |
| Salesperson | -35.425 | 998 | <.001 | <.001 | -1.171 | -1.24 | -1.11 |
| Hairdresser | -34.936 | 998 | <.001 | <.001 | -1.172 | -1.24 | -1.11 |
| Unemployed | -32.270 | 998 | <.001 | <.001 | -1.101 | -1.17 | -1.03 |
| Higher education | -33.090 | 998 | <.001 | <.001 | -1.124 | -1.19 | -1.06 |

*Note.* Negative t-values represent a negative deviation from the scale midpoint, whereas positive t-values represent positive deviations from the scale midpoint. Any mean differences are significant at the $p<.001$ level.

**APPENDIX D**

**SURVEY QUESTIONS ON LIFE AND BURGLARY INSURANCE**

*"With life insurance, you pay a higher premium if the insurer expects you to die earlier. Because if you die earlier, you will pay a premium for a shorter period. The insurer wants to charge higher premiums to people who are likely to die earlier. In the Netherlands, poorer people live on average seven years shorter than richer people. The insurer therefore charges higher premiums to poorer people. People with a certain income often live in certain neighborhoods.*

*The insurer charges higher prices in poorer neighborhoods."*

*"With burglary insurance, you pay a higher premium if the risk of burglary in your home is higher. Suppose poorer neighborhoods have more burglaries. An insurer charges higher premiums in neighborhoods with more burglaries. This leads to poorer people paying higher premiums on average."*